# Harnessing self-sensitised scintillation by supramolecular engineering of CsPbBr$_3$ nanocrystals in dense mesoporous template nanospheres

Xiaohe Zhou[1], Matteo L. Zaffalon[1,6], Emanuele Mazzola[2], Andrea Fratelli[1,7], Francesco Carulli[1], Chenger Wang[1], Mengda He[3], Francesco Bruni[1,6], Saptarshi Chakraborty[1], Leonardo Poletti[4], Francesca Rossi[4], Luca Gironi[2,6*], Francesco Meinardi[1], Liang Li[5] and Sergio Brovelli*[1,6]

[1]Dipartimento di Scienza dei Materiali, Università degli Studi di Milano-Bicocca, Via R. Cozzi 55, 20125, Milano, Italy

[2] Dipartimento di Fisica, Università degli Studi di Milano-Bicocca, Piazza della Scienza, 20125 Milan, Italy

[3]School of Environmental Science and Engineering, Shanghai Jiao Tong University, Shanghai 200240, China

[4]IMEM-CNR, Parco Area delle Scienze 37/A - 43100 Parma, Italy;

[5]Macao Institute of Materials Science and Engineering (MIMSE), Macau University of Science and Technology, Taipa 999078, Macao, China

[6]INFN - Sezione di Milano - Bicocca, Milano I-20126, Italy

[7]Nanochemistry, Istituto Italiano di Tecnologia, Via Morego 30, 16163, Genova, Italy

Email: Sergio Brovelli - sergio.brovelli@unimib.it, luca.gironi@unimib.it

* Corresponding author

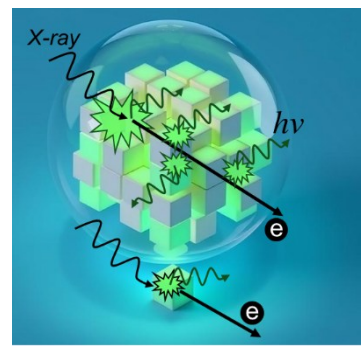

**Perovskite-based nanoscintillators, such as CsPbBr$_3$ nanocrystals (NCs), are emerging as promising candidates for ionizing radiation detection, thanks to their high emission efficiency, rapid response, and facile synthesis. However, their nanoscale dimensions — smaller than the mean free path of secondary carriers — and relatively low emitter density per unit volume, limited by their high molecular weight and reabsorption losses, restrict efficient secondary carrier conversion and hamper their practical deployment. In this work, we introduce a strategy to enhance scintillation performance by organizing NCs into densely packed domains within porous SiO$_2$ mesospheres (MSNs). This engineered architecture achieves up to a 40-fold increase in radioluminescence intensity compared to colloidal NCs, driven by improved retention and conversion of secondary charges, as corroborated by electron release measurements. This approach offers a promising pathway toward developing next-generation nanoscintillators with enhanced performance, with potential applications in high-energy physics, medical imaging, and space technologies.**

In recent years, the need to overcome the performance and scalability limitations of current ionising radiation detectors[1] has stimulated a growing interest in new generation scintillator materials that offer innovative platforms to address technological challenges in areas such as medical and high energy physics[2, 3], space exploration[4] and radiometric imaging[5]. These include so-called nanoscintillators[2, 6], which consist of chemically produced semiconductor nanoparticles that offer a combination of the high average atomic number Z required for interaction with ionising radiation[2, 7, 8] typical of expensive inorganic scintillator crystals, intense and ultrafast scintillation promoted by the formation of band-edge multi-excitonic states[8-10], radiation stability[11], and production scalability comparable to or superior to traditional aromatic chromophores used in plastic scintillators[12]. Among these, lead halide perovskite nanocrystals (NCs) of the general formula $CsPbX_3$ (where X is Cl, Br or I) have rapidly become the archetypal material[13], further favoured by their characteristic defect tolerance[14], wide chromatic tunability[15], radiation hardness[11] favoured by the mobility of the ionic lattice, which promotes self-healing[16], fast scintillation dynamics[17, 18] and the particularly advantageous ease of synthesis in large quantities at room temperature[12, 19]. An increasing number of studies have therefore been devoted to optimising their optical properties[20], their compatibility in plastic matrices[8, 9, 21, 22], including scintillating polymers[9, 23], and to obtain a detailed understanding of the mechanism of scintillation at the nanoscale, both at the level of isolated NCs[18, 24] and at the level of nanocomposites[9, 10, 25]. Based on these advances, important guidelines are now available for the design of $CsPbX_3$ NC-based scintillators that maximise the speed and efficiency of the scintillation process in both direct scintillators, which exploit the emission of NCs following ionizing excitation, and sensitised scintillators, in which NCs stimulate the luminescence of complementary organic chromophores.

However, important challenges remain in making this new class of nanocomposite scintillators applicable in real-world contexts. Indeed, a key aspect of nanocomposite scintillators is that the particle size is significantly smaller than the average free path of secondary charges (e.g. electrons) produced as a result of primary interactions with ionising radiation[25-27], which causes the electromagnetic shower produced by a NC to escape from the source particle[8, 18, 25]. This favours their use as sensitizers of secondary emitters through both their emission and their shower (still however lowering the scintillation speed) but limits the ability of matrix dispersed NCs to retain and convert energy into ultrafast light pulses. As in the case of conventional plastic scintillators, it is therefore important to design strategies also capable of the trapping and conversion of secondary charges. In nanocomposite scintillators based on heavy NCs, however, this need is met by a seemingly intrinsic limitation represented by their high Z, which gives average molecular weights of many orders of magnitude higher than those of conventional organic dyes (200-4000 kg/mmol for 3-8 nm NCs vs. 0.2-0.5 kg/mmol). In X-ray imaging screens, this is actually advantageous because it allows high stopping power from a few layers of NCs, minimising scattering while improving image quality[28], while in massive devices it means that for the same mass percentage, the number of NCs in a plastic matrix is orders of magnitude lower than in their organic counterparts, with a correspondingly greater interparticle spacing. Consequently, although nanocomposite scintillators exhibit scintillation efficiencies competitive with even commercial plastic scintillators when irradiated with soft radiation (favoured by their greater primary interaction capacity) [8, 22, 27], the use of high energy radiation becomes increasingly problematic precisely because of the increasing weight of secondary processes relative to primary ones as the energy of the incident radiation increases (progressively reducing the fraction of energy deposited within a single NC) and the difficulty of exploiting them effectively. From a practical point of view, achieving emitter densities, in terms of $NC/cm^3$, comparable to those of molecular scintillators (~1wt% corresponding to ~$10^{20}$ molecule/$cm^3$), by proportionally increasing the NC load, would entail weight fractions incompatible with the chemistry of real plastic nanocomposites (up to over 100% of the volume) and would entail high optical losses due to reabsorption and scattering of the scintillation light. As a result, effective collection and conversion of secondary carriers is an open challenge in NC-based scintillators.

In this context, recent results on polyvinyl toluene nanocomposites with high loading of $CsPbBr_3$ NCs showed a substantial increase in scintillation yield as a result of partial phase segregation of NCs in high particle density domains[23]. This local densification, albeit uncontrolled, resulted in a quadratic growth of scintillation intensity with NC loading compared to a linear growth of the same in similar solutions, leading to a 10-fold increase in scintillation yield at high loadings and suggesting the possibility of exploiting both direct intra-NC interactions and an inter-NC self-sensitisation process via the secondary shower by supramolecular control of the local spatial distribution of NC in the matrix. Such a scheme has not yet been demonstrated and might provide a viable strategy for advancing nanocomposite scintillators.

Here, we aim to contribute to this effort by investigating the effectiveness of engineering the arrangement of scintillating NCs into dense domains that locally mimic the inter-emitter spacing of organic scintillators, thus allowing to exploit the advantages of the high $Z$ of heavy NCs while also converting secondary charges (see sketch in **Figure 1a**). To this end, we have exploited the possibility of synthesising $CsPbBr_3$ NCs inside the pores of porous $SiO_2$ mesospheres (MSNs) [29, 30], which act both as a templating agent for the realisation of distinct and processable supramolecular units containing isolated NCs in a controlled number and distribution, and as a protective shell that stabilises the host particles against external agents ($H_2O$, environmental pollutants) and prevents the release of Pb ions into the environment[29, 31], potentially allowing their use also in biological contexts[32]. Radiometric experiments on model samples composed of NCs of the same size and specifically designed to exhibit identical optical properties in both the excitonic ($X$) and biexcitonic ($XX$) regimes show that, for the same total concentration of NCs in solution, their arrangement in mesospheres gives rise to a substantially greater radioluminescence (RL) than their colloidal counterparts, and that the intensity of the RL increases progressively with the size of the MSNs for the same density (NCs/cm$^3$) or their density at constant MSN size, resulting in up to a 40-fold increase in RL. Consistent with enhanced density of photoelectrons inside the MSNs following ionizing excitation, the increase in scintillation is accompanied by a progressive growth in the ultrafast contribution by charged excitons (or trions, $X^*$) to the emission kinetics, which is typically absent in colloidal suspensions or dilute composites[18, 22], opening to the possibility of further improving the timing capability of individual NCs by their engineered assembly. The direct correlation between the increase in scintillation properties and the conversion of secondary charges within the NC containing MSNs is further independently demonstrated by the experimental measurement of singlet oxygen ($^1O_2$) generation in aqueous solution prompted by the electric shower released by the particles following X-ray excitation, which anticorrelates quantitatively with the RL intensity trend. Finally, Monte Carlo simulations with Geant4 show that the amount of deposited energy increases proportionally to the degree of local densification and provide a deeper insight into the self-sensing mechanism as a function of material parameters. These results demonstrate a useful paradigm for the design of nanoscintillators and open up the possibility of realising metasolids that best exploit the advantages of dimensionally confined materials for radiation sensing, also potentially expanding the applicability of recent strategies for NCs embedding inside metal–organic frameworks (MOF)[30], amorphous network structures[33], and mesoporous nanoparticles[29, 34] that have been employed to preserve the optical properties in harsh environments.

**Results and Discussion**

The scintillation properties of $CsPbBr_3$ NCs are determined by structural and photophysical parameters that jointly determine the behaviour in the *X* and *XX* regimes, both of which are present under ionising excitation[9, 18]. In particular, size determines the average energy deposited in a NC and hence the average excitonic occupancy (*N*) produced following a multi-exciton generation process[2, 8, 18, 35] and, via volume scaling[36], the efficiency of the Auger recombination process, which determines the *XX* and *X\** recombination yield. The PL quantum yield ($\Phi_{PL}$) determines the contribution of *X* (generated directly or as a result of *XX* recombination) to the total emission. The relationship between scintillation intensity and dynamics and these photophysical parameters follows a complex dependence[18], which weakens a performance comparison by a posteriori corrections of experimental data. For this reason, and in order to accurately assess the local densification effects of NCs, $CsPbBr_3$ NC-based samples with nearly identical optical and structural properties were realised. First, we synthesised a series of uniform MSNs following a modified protocol, which were then used as nanotemplates for the confined growth of $CsPbBr_3$ NCs[30]. The MSNs exhibited uniform sizes of *d*=100, 200 and 350 nm (**Figure S1**) with long-range ordered pore structures with comparable pore sizes and large specific surface areas. The templates were then soaked in the solution of the precursor salts ($CsBr$ and $PbBr_2$) and then dried at 80 °C to synthesise $CsPbBr_3$ NCs in their inner pores. The resulting mixture was placed in a furnace and heated to 600 °C. The cooled samples were washed with ultrapure water and dried to obtain the final product. In order to study the effect of the size of artificially densified NC domains, a set of NC loaded MSNs (hereafter indicated as NC-MSNs) of the three different sizes was prepared by keeping the density of $CsPbBr_3$ NCs per MSN constant at 10 wt%, corresponding to $\rho$ = 77, 660 and 2100 NCs/MSN respectively. As shown in **Table S1**, which reports the results of the chemical analysis of the samples, the actual $CsPbBr_3$ content of the *d*=350 nm NC-MSNs was lower than the nominally expected value ($\rho$ = 3300 NCs/MSN), probably due to the difficulty of the reagents to reach the innermost particle pores, resulting in impregnation of only the outermost nanosphere shell. However, this sample was kept in the set throughout the study because, as shown below, its off-trend behaviour in both radioluminescence (RL) yield and electron retention capability actually indirectly confirms the size dependence of the others (vide infra). For the set of samples with increasing density of NCs, *d*=200 nm MSNs were used as templates and the loading of NC concentration was progressively increased from 10wt% (comparable to the *d*=100 nm NC-MSNs) to 20wt% and 30wt% to obtain densities of $\rho$ = 1200 and 1800 NC/MSNs. Finally, a sample of colloidal $CsPbBr_3$ NCs of comparable size (8 ± 0.9 nm) was prepared according to a literature protocol[37].

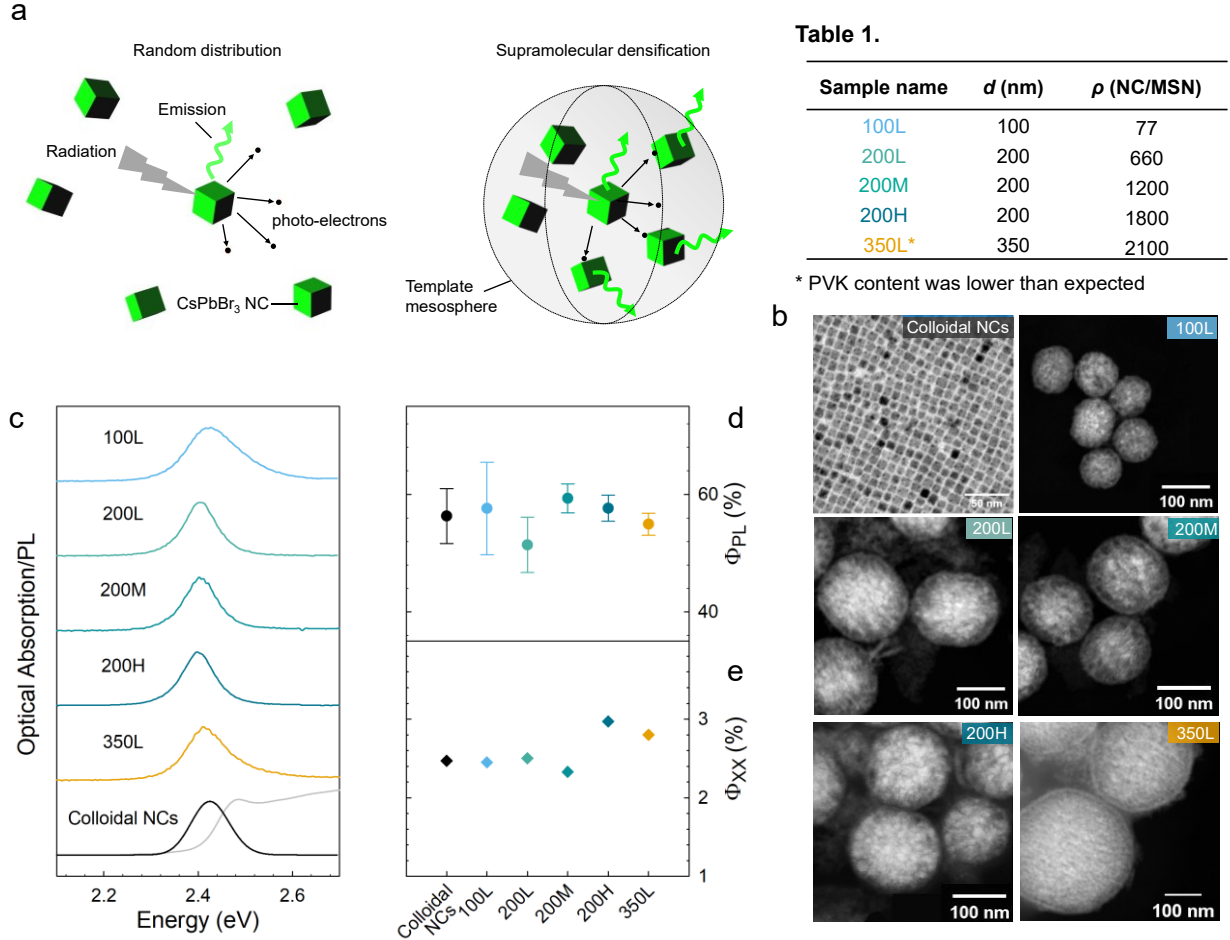

**Figure 1. Concept of locally densified NC scintillators and controlled model system. a.** Scintillation mechanism in isolated NCs vs. engineered densified structures. **b**, TEM images of NC-MSNs of sizes $d$ = 100nm, 200nm and 350nm. **c**. Optical absorption (grey line), PL spectra of colloidal NC (black line) and NC-MSN (colored lines) dispersed in toluene. Excitation energy $E_{ex}$=3.05eV. **d**, PL quantum yield ($\varphi_{PL}$) and **e** biexciton quantum yield ($\varphi_{XX}$) of colloidal NCs and NC-MSNs. $\varphi_{XX}$ is extracted from the fitted single exciton and biexciton components of TA dynamics at their corresponding 1S bleach maximum using successive subtraction method[1].

**Figure 1b** shows the transmission electron microscopy images of the investigated NC-MSNs together with the corresponding colloidal suspension, showing distinct crystalline $CsPbBr_3$ NCs with a cubic structure (as confirmed by X-ray diffraction measurements, **Figure S2**) and an average size of 8±0.9 nm within the MSNs. The PL spectra of the samples (**Figure 1c**) show the typical excitonic emission peak of $CsPbBr_3$ NCs at about 2.4 eV, confirming the comparable size of the embedded NCs. The PL quantum yield ($\Phi_{PL}$) of all systems dispersed in toluene was measured in an integrating sphere and shows an essentially constant value within the error bars of about 55±6% (**Figure 1d**). To investigate the local optical properties of the NC-MSNs, cathodoluminescence (CL) measurements were performed on all samples. **Figure S3** shows CL maps acquired using an electron beam current of 300 pA and an accelerating voltage of 5 kV. The corresponding spectra are shown in **Figure S4**; histograms showing the CL intensity per unit area of the analysed particles are reported in **Figure S5**. For the spheres with diameters $d$ = 100 nm, the number of NCs in each sphere (77 as indicated in Table 1) is too low to allow successful CL mapping. CL images of the $d$ = 200 nm samples at increasing concentrations clearly show a corresponding increase in luminescence, particularly between the 200L and 200M samples. Regarding the 350 nm MSNs, a significant inhomogeneity in NC incorporation and a reduction in CL mean intensity compared to the 200 nm MSNs was observed. This is consistent with the decrease in RL intensity reported in Figure 2.

As mentioned above, the scintillation mechanism in NCs is strongly dependent on the multi-exciton photophysics. Therefore, we performed transient transmission measurements as a function of excitation fluence to assess the *XX* yield. This was done following the standard approach in carrier dynamics studies in NCs[38], i.e. by measuring the beach dynamics of the 1S absorption feature at increasing fluence (see **Figure S6, S7**) and subtracting the lowest fluence dynamics due to *X* decay from the other to extract the *XX* decay rate, $k_{XX}$. The *XX* efficiency was then estimated under the common assumption that the low fluence dynamics correspond to *X* radiative decay (with rate $k_X^R$) and considering the statistic relation $k_{XX}^R = 4k_X^R$, so that the *XX* yield can be written as $\Phi_{XX} = k_{XX}^R/k_{XX}$.[38] The resulting $\Phi_{XX}$ values for the whole sample set are shown in **Figure 1e** and again show a constant value of about 2.5%, in agreement with previous reports on CsPbBr$_3$ NCs of comparable size[39].

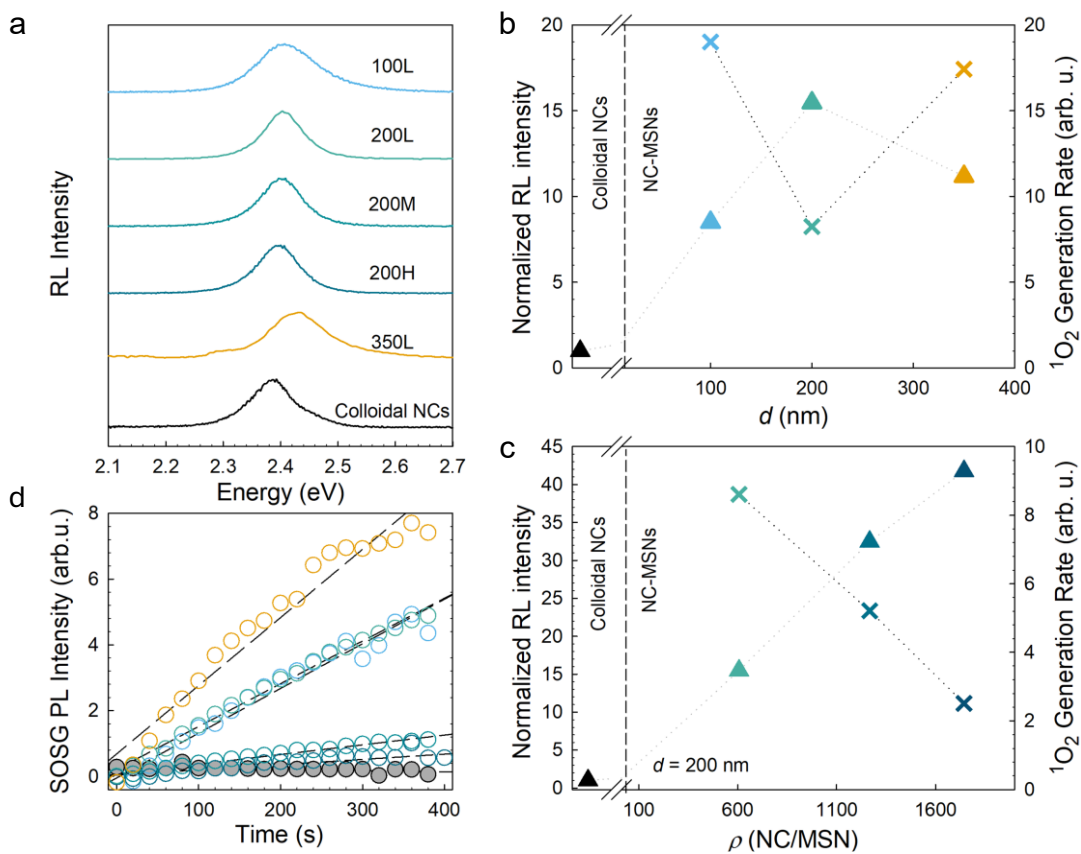

**Figure 2. Radioluminescence and electron release from supramolecularly densified nanoscintillators. a**, RL spectra of NC-MSNs (colored lines) and analogous colloidal CsPbBr$_3$ NCs reference (black line) dispersed in dodecane. The total CsPbBr$_3$ contents in each sample solution are maintained at 0.1wt%. Same color code is applied to all panels. **b**, Integrated RL intensity of NC-MSN normalized to colloidal NCs, overlayed with $^1$O$_2$ generation rate as an indicator of released electrons as a function of increasing MSN size (constant density as shown in **Table 1**) **c**, RL/$^1$O$_2$ generation rate as a function of number of NCs per MSN ($d$=200 nm). The $^1$O$_2$ generation rates are extracted from the slope of the respective growth of SOSG intensity (excited at 473nm) normalized by the initial SOSG intensity of pristine MSN (black circles) and NC-MSNs (colored circles) with increasing X-ray exposure time shown in '**d**'.

After confirming the validity of the model sample set and the consistency of the respective optical properties as the degree of local organization of the NCs was varied, we proceeded with a quantitative comparison of their scintillation under X-ray excitation from a tungsten cathode (average energy 10 keV, 20 mA). For this purpose, dispersions with the same average concentration of $CsPbBr_3$ NCs were used, as confirmed by optical absorption measurements (**Figure S8**). This implies that the content of emitting species is constant (i.e., $CsPbBr_3$ NCs), while the number of NC-MSNs in solution changes on a case-by-case basis depending on their NCs content. To eliminate possible artifacts due to different outcoupling of the scintillation light and to avoid spurious contributions from the solvent, all NC-MSNs samples were surface functionalized with octadecyltrichlorosilane (OTS) to maximize their solubility and redispersed in octane, which has no scintillation of its own[2]. Colloidal NCs did not require surface functionalization. Finally, to show differences due to local densification of the NCs by the template, the total concentration of $CsPbBr_3$ was kept low so that inter-NC interactions in the colloidal solution were negligible. The RL spectra are shown in **Figure 2a** and show agreement with the corresponding PL and CL, confirming the absence of parasite emission due to MSNs or defect states. Given the complete similarity of the spectral features, the corresponding scintillation intensities measured under identical excitation and collection conditions show substantial differences, confirming the hypothesized self-sensitization effect due to the spatial organization imposed by the MSN templates. Specifically, **Figure 2b** shows the RL intensity normalized to the value for the colloidal solution of NC-MSNs with increasing size. Although the concentration of $CsPbBr_3$ NCs in solution was constant, increasing the size of the mesosphere gradually enhanced the RL intensity, reaching a 15-fold enhancement in the $d$=200 nm NC-MSNs compared to the respective colloidal solution. We note that the use of larger template MSNs ($d$=350 nm) led to a decrease in RL intensity, consistent with the lower atomic content shown in **Table S1**, resulting in greater electronic losses. This is further confirmed by the electron release measurements and time resolved RL experiments discussed below (vide infra). Increasing the NC density within the $d$=200 nm MSNs - for which effective embedding of large amounts of $CsPbBr_3$ was possible - resulted in further progressively more intense RL, yielding a total enhancement of 40 times with respect to comparable amount of colloidal NCs (**Figure 2c**). The dramatic enhancement of RL in NC-MSNs agrees well with recent theoretical results by Villa et al. that revealed nearly complete deposition of energy inside $SiO_2$ nanospheres following X-ray excitation, which, in the current case results in stronger NC excitation[40]. We emphasize that the optical density due to $CsPbBr_3$ NCs is constant for all samples studied, and thus these experiments validate the concept of engineering the local distribution of NCs to enhance their scintillation yield without introducing additional loss due to reabsorption that would be direct consequence of enhancing the average NC concentration, which is currently an unresolved limitation in the use of NCs with excitonic emission in scintillation[23]. As a result of such a dramatic improvement in RL with respect to colloidal NCs, the LY of 300 μL (1%wt in a 0.5 cm high cylindrical cuvette) of the NC-MSN suspension measured in the same excitation and collection geometry at room temperature as a commercial plastic scintillator EJ-276D (LY = 8600 photons/MeV)[41] of the same volume (in both cases resulting in essentially complete deposition of the incident X-ray excitation) is ~40'000±1000 photons/MeV, which is also consistent with the observed enhancement with respect to reported results on diluted CsPbBr$_3$ NCs solutions and composites[18, 23]. However, we emphasize that the LY of NC-based scintillators, especially in the light of the present results, must be treated with great caution, since it strongly depends on their concentration and on their ability to convert secondary carriers, which in turn is a decreasing function of the irradiation energy. Therefore, in general and also in the present case, a significant nonlinearity of the LY with the irradiation energy should be expected.

To strengthen the attribution of the observed increase in RL intensity to increased conversion of secondary charges into scintillation light, we independently measured electron release following ionizing interaction. To do this, we used a fluorescent singlet oxygen sensor (SOSG)[42], commonly used in radiotherapy studies, to monitor in situ the evolution of $^1O_2$ generated following electron release by X-ray sensitizers under irradiation. In its non-oxidized form, SOSG is non-emissive, while its endoperoxide derivative, formed as a result of oxidation by $^1O_2$, exhibits a PL at 530 nm (**Figure S9**), the intensity of which is commonly used to quantify the concentration of $^1O_2$ during X-ray irradiation, which in turn is proportional to the release of electrons into solution. In this study SOSG was mixed with NC-MSN in aqueous solution and excited by a laser at 473 nm; each scan lasted 10 minutes of continuous exposure from the same X-ray source used for the RL measurements. **Figure 2d** reports the time evolution of the SOSG PL intensity using NC-MSNs as well as empty MSN templates and bare water solution. Consistent with previous results[42], the introduction of heavy NC-MSNs largely increases the SOSG emission, which further suggests their potential use as luminescent X-ray sensitizers in radiotherapy. Most relevant to this work, the $^1O_2$ generation rates, extracted as the slope of the curves in **Figure 2d** and reported as crosses in **Figure 2b** and **2c**, anticorrelate remarkably well with the respective RL intensities, thus independently supporting the picture that the stronger RL intensities are in fact associated to better retention and conversion of secondary electrons, once again in agreement with previous reports[40].

The local organization of NCs into MSNs, in addition to increasing the scintillation yield, also has a relevant impact on the RL kinetics, providing further independent evidence for the increased electron density within the NC-MSNs. As demonstrated in **Figure 3a**, which compares the RL time decays under X-pulsed excitation of colloidal NCs and NC-MSNs ($d$=200 nm, normalized to their long time tails to emphasize the different ultrafast contributions, the whole set of data is reported on **Figure S10**), the decay of colloidal NCs is dominated by a 10 ns component attributed to $X$ decay and a relatively small ultrafast contribution due to the combination of $X^*$ and $XX$ emission[18]. In accordance with the literature, the decay profile was fitted by deconvolution with the instrumental response (see Methods), yielding three kinetic components with $\tau_X$ = 10 ns, $\tau_{X^*}$ = 1 ns and $\tau_{XX}$ = 80 ps as shown in **Figure 3b** (top panel), which reports the lifetime values and respective relative weights. The relative weights of the $XX$ and $X^*$ components are consistent with CsPbBr$_3$ NCs of the same size[39], with the latter being nearly negligible, confirming the essentially total release of secondary carriers outside the primary NC and the impossibility of reaching (and thus charging) another NC in dilute solution. The average excitonic population extracted from the time resolved trace of the colloidal solution according to ref. [38] is N~1, which agrees with recent results on NCs of the same size under the same X-ray excitation[18].

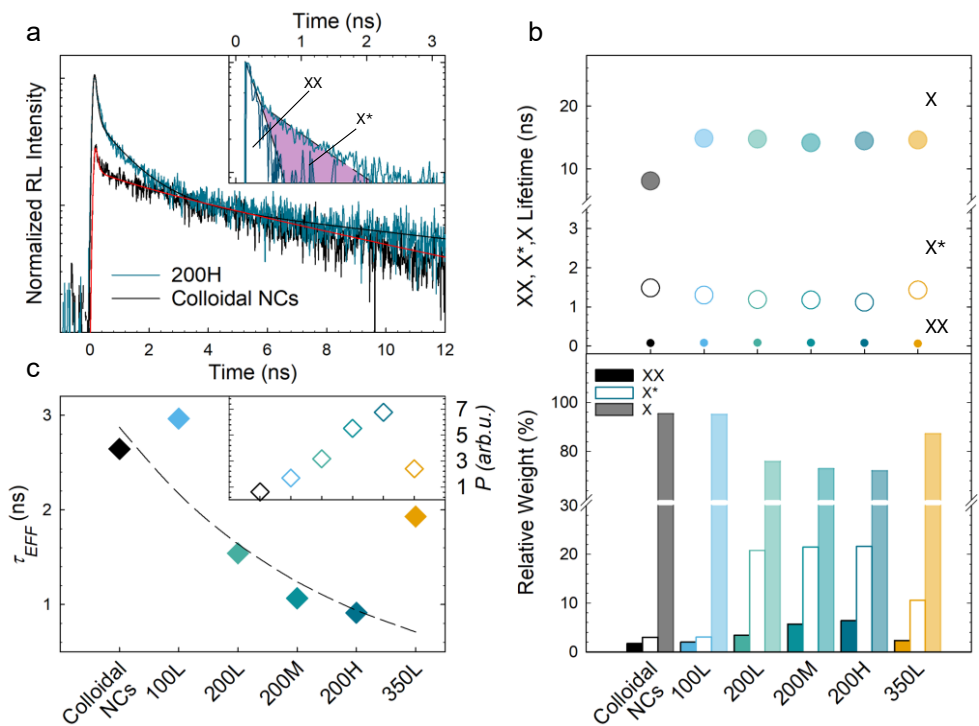

**Figure 3**. **Dynamic effects of supramolecular densification. a**, Time-resolved RL of colloidal NCs (black line) and 200H (dark green line), normalized to single exciton tail. Inset: X* (colored line) and XX(black line) dynamics of 200H are explored at fast-time scale (time window 0-3ns) by subtracting contribution from single exciton showing charged exciton following successive subtraction method[1]. The shaded area highlights the contribution from X* in the NC-MSN. **b**. Decay lifetimes (upper panel) and their relative weight (lower panel) of X* (hollow) and XX (solid) of colloidal NCs and MSN samples. **c.** Effective lifetime $\tau_{EFF}$ of colloidal NCs and MSN. Inset: Pulse quality factor $P$ across the sample set following the same color code as the main plot.

More importantly, NC-MSNs show a significantly more intense ultrafast contribution, driven by a marked increase in the $X^*$ component (~1000% increase, **Figure 3b**), with a slight increase in the $XX$ contribution. This is consistent with the increased density of photoelectrons in NC-MSNs, which, in addition to exciting the luminescence of NCs, promotes the formation of charged states. As a result, the RL becomes more intense and faster, reaching an effective lifetime $\tau_{EFF} \sim 1\ ns$ in the highly loaded d=200 nm particles (sample 200H, $\tau_{EFF}$ is defined as the weighted harmonic average of the decay contributions, see Methods). As a counter-evidence to this effect, the dynamics of the $d$=350 nm NC-MSNs (sample 350L) with lower RL and higher $^1O_2$ generation rate shows a relative weight of the $X^*$ component lower than the corresponding $d$=200 nm NC-MSNs (**Figure 3b**). Finally, it is instructive to evaluate the overall trend of the relative performance in pulsed detection resulting from the RL intensification ($I_{RL}$) and the concomitant reduction of $\tau_{EFF}$ in NC-MSN under X-ray excitation by the so-called pulse quality[43], expressed as $P \sim \sqrt{\frac{I_{RL}}{\tau_{EFF}}}$. Considering $I_{RL}$ = 1 for colloidal NCs and the relative RL intensities shown in **Figure 2b,c**, an improvement in $P$ of up to 9 times from the colloidal NCs to the 200L NC-MSNs is obtained (inset of **Figure 3c**), which is valuable for fast timing applications.[8, 22]

With the aim to gain a deeper understanding of the energy deposition processes in CsPbBr$_3$ NCs confined within MSNs, a simulation was finally developed using Geant4 (version 11.1.2), a toolkit for simulating radiation-matter interactions. Two simulation codes were implemented to model different geometric configurations that replicate the experimental conditions: a single NC-MSN system (code A) and a larger ensemble containing multiple NC-MSNs (code B). In code A, the geometry consists of a single SiO$_2$ nanosphere with a diameter of 200 nm, filled with randomly distributed, non-overlapping CsPbBr$_3$ NCs (each with an edge length of 8 nm, **Figure 4a**). Code B, by contrast, models a system in which multiple NC-MSNs (as defined in code A) are randomly distributed — without overlap — within a 3 μm-sided polyacrylate cube (**Figure 4b**). The cube size was selected to balance computational cost and the number of NC-MSNs. To accurately simulate electromagnetic interactions, the *G4EmStandardPhysics_option4* physics list was used, chosen for its high precision in electron tracking and use of advanced low-energy models. In both codes A and B, the radiation source was modelled as an X-ray beam (10$^7$ photons) with a continuous energy spectrum ranging from 0 to 20 keV, matching the experimental RL conditions.

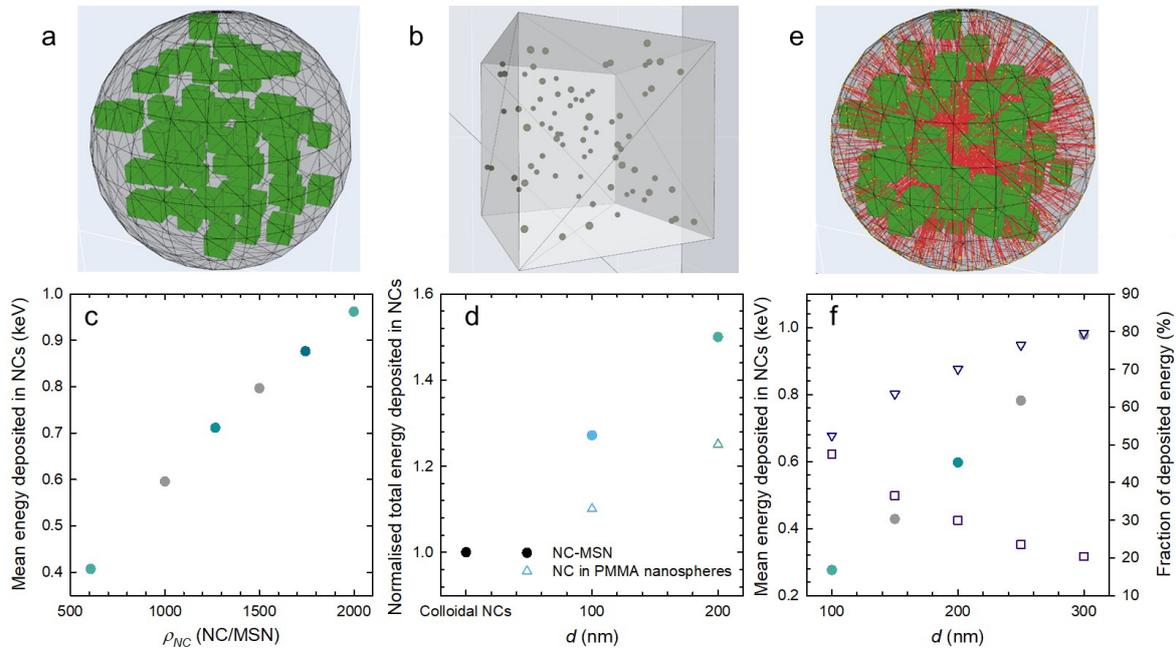

**Figure 4. Monte Carlo simulations of energy deposition. a**, Schematic representation of a single NC-MSN containing 77 CsPbBr$_3$ nanocrystals, generated using code A. **b**, Simulation of a polyacrylate cube (side length: 3 μm) embedding 77 NC-MSNs, performed using code B. **c**, Mean energy deposited per X-ray in CsPbBr$_3$ NCs as a function of increasing nanocrystal density ($\rho_{NC}$) within a single NC-MSN, simulated with code A. **d**, Total energy deposited in CsPbBr$_3$ NCs, expressed as a percentage relative to the colloidal NC case, as a function of increasing MSN diameter. Grey data points correspond to simulations in which SiO$_2$ was replaced by polymethyl methacrylate. These simulations were conducted using code B. **e**, Geometry of a single NC-MSN identical to that in panel **a**, featuring an electron source located within a central NC. **f,** Mean energy deposited per electron in NC-MSNs of varying diameters (circles). Squared indicate the fraction of energy deposited in the NC where the electrons are initially generated, while triangles represent the fraction deposited in surrounding NCs. Values are normalized to the total energy deposited in all other NCs.

Code A was used to investigate how energy deposition in CsPbBr$_3$ NCs varies with increasing NC density ($\rho_{NC}$) within a single NC-MSN. As shown in **Figure 4c**, the energy deposited per X-ray increases linearly with $\rho_{NC}$. To further examine how the spatial arrangement of NCs within MSNs affects total energy deposition, code B was used to simulate three distinct configurations of NC ensembles embedded in a polyacrylate matrix: *i*) 5929 randomly dispersed individual CsPbBr$_3$ NCs; *ii*) 77 NC-MSNs (100 nm diameter), each containing 77 NCs; *iii*) 10 NC-MSNs (200 nm diameter), each containing 616 NCs. Finally, to isolate the respective contributions of geometric confinement and material composition to the observed enhancement in energy deposition, a conceptual simulation (*gedankenexperiment*) was performed in which the SiO$_2$ in the MSNs was replaced with polymethyl methacrylate. The results (**Figure 4d**) show an increase in energy deposition when NCs are confined within MSNs. This enhancement appears to arise equally from geometric confinement and the higher atomic number of the MSN material. We point out that the lack of quantitative agreement between the experimental RL enhancement and the simulated energy deposition in **Figure 4d** is likely to be due to scintillation effects, which are not captured by the current simulation and will be addressed in a dedicated study due to their complexity. Finally, we investigated the energy deposition from secondary electrons in NCs embedded within MSNs of varying diameters using code A. For this simulation, a source of $10^6$ electrons was randomly generated within a single NC located at the center of the MSN (**Figure 4e**). This setup was further used to evaluate how the deposited energy is distributed between the NC where the electrons originate and the surrounding NCs. As illustrated in **Figure 4f**, the total energy deposited within the system increases with the diameter of the MSN. In smaller NC-MSNs, a substantial portion - nearly 50% - of the energy is deposited in the NC where the electrons are initially generated. As the MSN size increases, the relative contribution of the source NC decreases, indicating that while overall energy retention becomes stronger, the energy deposition is more evenly distributed among surrounding NCs.

In conclusion, this study demonstrates that the local densification of CsPbBr$_3$ NCs within porous silica mesospheres enables a remarkable enhancement of scintillation properties up to 40 times greater than those of equivalent colloidal NC solutions. This improvement is attributed to the increased efficiency in converting secondary charges generated by the interaction with ionizing radiation. The results highlight how the spatial organization of NCs into high-density domains can overcome the intrinsic limitations of traditional nanocomposites, paving the way for the development of high-performance "metasolid" scintillators.

**Acknowledgments**

This work was funded by Horizon Europe EIC Pathfinder program through project 101098649 – UNICORN, by the European Union -Next Generation EU, Mission 4 Component 1 CUP H53D23004670006, and through the Italian Ministry of University and Research under PNRR—M4C2-I1.3 Project PE_00000019 "HEAL ITALIA". This research is funded and supervised by the Italian Space Agency (Agenzia Spaziale Italiana, ASI) in the framework of the Research Day "Giornate della Ricerca Spaziale" initiative through the contract ASI N. 2023-4-U.0t

**Competing interests**

The authors declare no competing interests.

# Supporting Information

# Harnessing self-sensitized scintillation by supramolecular engineering of CsPbBr$_3$ nanocrystals in dense mesoporous template nanospheres

Xiaohe Zhou[1], Matteo L. Zaffalon[1,6], Emanuele Mazzola[2], Andrea Fratelli[1,7], Francesco Carulli[1], Chenger Wang[1], Mengda He[3], Francesco Bruni[1,6], Saptarshi Chakraborty[1], Leonardo Poletti[4], Francesca Rossi[4], Luca Gironi[2,6*], Francesco Meinardi[1], Liang Li[5] and Sergio Brovelli*[1,6]

[1]Dipartimento di Scienza dei Materiali, Università degli Studi di Milano-Bicocca, Via R. Cozzi 55, 20125, Milano, Italy
[2] Dipartimento di Fisica, Università degli Studi di Milano-Bicocca, Piazza della Scienza, 20125 Milan, Italy
[3]School of Environmental Science and Engineering, Shanghai Jiao Tong University, Shanghai 200240, China
[4]IMEM-CNR, Parco Area delle Scienze 37/A - 43100 Parma, Italy;
[5]Macao Institute of Materials Science and Engineering (MIMSE), Macau University of Science and Technology, Taipa 999078, Macao, China
[6]INFN - Sezione di Milano - Bicocca, Milano I-20126, Italy
[7]Nanochemistry, Istituto Italiano di Tecnologia, Via Morego 30, 16163, Genova, Italy
Email: Sergio Brovelli - sergio.brovelli@unimib.it, luca.gironi@unimib.it
* Corresponding author

## Methods
### Synthesis of CsPbBr$_3$ nanocrystals.

*Chemicals.* Cesium carbonate (Cs$_2$CO$_3$, 99.9%, Sigma-Aldrich), diisooctylphosphinic acid (DOPA, 90%, Sigma-Aldrich), lead bromide (PbBr$_2$, 99.999%, Sigma-Aldrich), trioctylphosphine oxide (TOPO, 90%, Strem), n-octane (97%, Acros Organics), hexane (96%, Scharlau), Oleic acid (OA, 90%, Sigma-Aldrich), lecithin (≥97%, from soy, Roth); mesitylene (99%, Thermo Scientific Chemicals); phosphorus(V) oxychloride (99%, Sigma-Aldrich); 2-aminoethan-1-ol (≥99.0%, Sigma-Aldrich); acetic acid (>99.8%, Sigma-Aldrich); triethylamine (99%, Sigma-Aldrich); 2-octyl-1-dodecanol (97%, Sigma-Aldrich).

*Stock solutions.* PbBr$_2$-TOPO stock solution was prepared by dissolving 367 mg PbBr$_2$ and 2.15 g TOPO in 5 ml n-octane at 120 °C, followed by cooling down and dilution with 20 ml hexane. Cs-DOPA stock solution was prepared by reacting 100 mg Cs$_2$CO$_3$ with 1 ml DOPA in 2 ml n-octane at 120 °C, followed by cooling down and dilution with 27 ml hexane. For mixed-halide NCs, ZnCl$_2$-TOPO and ZnI$_2$-TOPO stock solution for anion exchange were prepared by dissolving 136 mg ZnCl$_2$ or 319 mg ZnI$_2$ with 1.933 g TOPO in 2.5 ml octane at 120 °C, followed by cooling down and dilution with 7.5 ml hexane. 2-ammonioethyl 2-octyl-1-dodecyl phosphate (C$_8$C$_{12}$-PEA) ligand was synthesized as reported in Ref. *Nature* 626, 542–548 (2024).

*Synthesis.* NCs were synthesized according to the modified PbBr$_2$-TOPO method in Ref. *Science* 377, 1406-1412 (2022). A 25 ml flask was loaded with 3 ml hexane and 2 ml PbBr$_2$-TOPO stock solution and stirred at 1200 rpm. Next, 1 ml Cs-DOPA stock solution was injected, and the solution was stirred for 3 minutes before the ligand was added (C$_8$C$_{12}$-PEA, 0.26 ml, 0.1M in mesitylene). The crude solution was rotary evaporated at room temperature until 5-7ml left, washed with acetone (0.4 eq.), centrifuged at 12100 rpm, and redispersed in n-octane. To obtain mixed-halide NCs, ZnCl$_2$-TOPO or ZnI$_2$-TOPO stock solutions were added (equimolar to the PbBr$_2$) before the washing, followed by stirring for a few minutes.

### Synthesis CsPbBr$_3$-SiO$_2$ MSN.

*Chemicals.* Cesium bromide (CsBr, 99.5%, Aladdin), lead bromide (PbBr$_2$, 99%, Aladdin), Potassium carbonate (K$_2$CO$_3$, 99%, Aladdin), cetyltrimethylammonium bromide (CTAB, 95%, Aladdin) and tetraethylorthosilicate (TEOS, 95%, Aladdin), sodium hydroxide (NaOH, 98%, Aladdin), cesium carbonate (Cs$_2$CO$_3$, 99.9%, Aladdin), 1-octadecene (ODE, 90%, Aladdin), oleylamine (OAm, 90%, Aladdin), oleic acid

(OA, 90%, Aldrich), F127 ($EO_{106}PO_{60}EO_{106}$, AR, Macklin), ethanol (99.5%, Sinopharm Chemical Reagent), methyl acetate (98%, Sinopharm Chemical Reagent), toluene (99.5%, Sinopharm Chemical Reagent), meso-tetra(4-sulfonatophenyl) porphyrin ($H_2TPPS^{4-}$) (Combi-Blocks).

*Preparation of MSNs.* MSNs were synthesized by the following procedure: cetyltrimethylammonium bromide (CTAB, 1.00 g) and F127 (100mg) was first dissolved in 480 mL of ultrapure water. Then, 3.0 mL of a NaOH (2M) solution was added and stirred in a water bath at 80 °C for 30min. 5.0 mL TEOS was then dropped into the above solution. The mixture was stirred for 2 hours and the solid was collected by centrifugation, washed with ultrapure water, and dried at 80°C. To prepare the final MSNs the synthesized solid was calcined for 6 hours in an oxidizing atmosphere at 550 °C to remove the template. By changing the volume of the NaOH (2M) solution to adjust the size of MSNs, the MSNs of 100nm and 350nm diameter correspond to 2.1mL and 3.9mL of NaOH solution, respectively.

*NC-MSNs synthesis.* The preparation of NC-MSNs was performed following a previously optimized procedure. Specifically, for medium NC loading, 0.6 mmol of salt precursors (127.69 mg of CsBr and 220.20 mg of $PbBr_2$) were dissolved in 50 mL ultrapure water, sonicate for 5 min and stir continuously at 80 °C for 30 min until clear. Then around 1050 mg of MSNs (pre-dispersed in 20 mL ultrapure water) (the mass ratio of NC precursors: MSNs=1:3) was added into the above solution. The mixture is stirred continuously at 80°C until dry. The collected mixture was ground and calcined at 600°C temperature for 30 min with a heating rate of 5 °C/min in a muffle furnace under air atmosphere. After cooling to room temperature, the sample was ground and washed with ultrapure water several times to remove external NC or other salts. Finally, the washed sample was obtained by centrifugation and drying at 80 °C. By changing the mass ratio of NC precursors to MSNs to adjust the loading of NC, low loading and high loading correspond to NC precursors : MSNs= 0.5:3 and 2:3, respectively.

*Morphological and elementary characterization.* The powder X-ray diffraction (XRD) patterns of samples were performed by a Bruker D8 Advance X-ray Diffractometer at 40 kV and 30 mA using Cu Kα radiation (λ = 1.5406 Å). TEM/STEM imaging and EDX spectroscopy were performed in a JEOL JEM-2200FS microscope, operated at 200 kV, equipped with an high-angle annular dark field detector for Z-contrast imaging and an Oxford Xplore detector for compositional analysis. The particles were deposited by drop-casting on Cu grids with ultrathin carbon support film.

*Optical spectroscopy.* Optical absorption measurements were measured in octane with an Agilent Cary 60 UV–Vis spectrophotometer. PL measurements were performed by exciting the samples with a 405 nm pulsed diode laser (Edinburgh Inst. EPL 405, 40 ps pulse width), and collected with a TM-C10083CA Hamamatsu Mini-Spectrometer. PLQY for every sample was obtained by comparison with a standard with the same absorbance at the excitation energy. Time-resolved PL were carried out using the same 405nm source (Edinburgh Inst. EPL 405, 40 ps pulse width) in PL measurement; the emitted light was collected with a phototube coupled to a Cornerstone 260 1/4 m VIS-NIR Monochromator (ORIEL) and a time-correlated single-photon counting unit (time resolution ~400 ps). Ultrafast transient absorption spectroscopy measurements were performed on Ultrafast Systems Helios TA spectrometer. The laser source was a 10 W Ytterbium amplified laser operated at 1.875 kHz producing ~260 fs pulses at 1030 nm and coupled with an independently tunable optical parametric amplifier from the same supplier that produced the excitation pulses at 3.1 eV (400 nm). After passing the pump beam through a synchronous chopper phase-locked to the pulse train (0.938 kHz, blocking every other pump pulse), the pump fluence on the sample was modulated using a variable ND filter. The probe beam was a white light supercontinuum.

*RL measurements* Unfiltered X-rays were produced using a Philips PW2274 X-ray tube with a tungsten target, equipped with a beryllium window and operated at 20 kV to produce a continuous X-ray spectrum through bremsstrahlung. The scintillation light was detected using a liquid-nitrogen-cooled, back-illuminated, UV-enhanced CCD detector (Jobin Yvon Symphony II), coupled to a monochromator (Jobin Yvon Triax 180) with a 100 lines/mm grating. CL maps were acquired using an electron beam current of 300 pA and an accelerating voltage of 5 kV.

*LY measurements* Light yield values were determined by comparing the integrated RL intensity under 20 kV X-ray excitation ($\langle E \rangle \sim 7$ keV) with identical experimental conditions for a 1 wt% octane solution of CsPbBr$_3$ NCs and NC-MSNs placed in a 5 mm long crucible and a commercial EJ-276D plastic scintillator (LY = 8600 photons/MeV) of the same size and geometry used as a reference. In both cases, the sample size was chosen to ensure complete attenuation of the excitation beam.

*Time resolved scintillation measurements* The time-resolved RL was measured using a pulsed X-ray source consisting of a 405 nm ~70-ps pulsed laser hitting the photocathode of an X-ray tube (N5084, Hamamatsu) set at 40 kV. The emitted scintillation light was collected using an FLS980 spectrometer (Edinburgh Instruments) coupled to a PicoHarp 300 hybrid photomultiplier tube operating in TCSPC mode. The RL decay curves were analyzed using a least-squares fitting approach with the following formula, which accounts for the convolution with the instrument response function (IRF):

$$F(t) = IRF(t) \otimes \left( H(t - t_0) \cdot \left[ \sum_{i=1}^{2} a_i \cdot e^{-t/\tau_i} \right] \right) + C$$

where $t_0$ corresponds to the start of the emission process, *C* is the electronic background noise floor, and *H* is the Heaviside function. The experimental IRF was well described by a Gaussian profile (FWHM = 120 ps), and the weight of each component ($w_i$) was calculated as the integral of each convoluted function over the entire time window. The average lifetime was calculated using the re-normalized ratio of all components $\tau_i$ according to:

$$\tau_{eff} = \left( \frac{\tau_1}{w_{1n}} + \frac{\tau_2}{w_{2n}} \right)^{-1}, \qquad w_{in} = \frac{w_i}{w_1 + w_2}$$

*Singlet Oxygen Generation Measurement.* The optical probe SOSG has been purchased from Thermo Fisher. The SOSG powder has been diluted in a 1:10 solution of dimethyl sulfoxide (DMSO) and PBS, which has been used to disperse the CsPbBr$_3$-SiO$_2$ with a concentration of 4 mg/mL in water. The intensity of the SOSG fluorescence, which is directly proportional to the concentration of singlet oxygen in the environment, has been monitored during the X-ray exposure under continuous-wavelength laser light excitation at 473 nm.

*Evaluation of ROS production rate.* A 4 mL PBS solution containing the same concentration of SOSG used in the ROS production experiment (8.3×10$^{-5}$ M) was prepared and 0.5 mg of meso-tetra(4-sulfonatophenyl) porphyrin (H$_2$TPPS$^{4-}$), an efficient photosensitizer for singlet oxygen generation, were added and the final solution and kept under stirring in dark condition until the complete dissolution of H$_2$TPPS$^{4-}$. We performed the complete oxidation of SOSG via the photo-sensitizer approach rather than the radio-sensitizer to avoid exposing the sample under extremely high X-ray dose rate, which may result in undesired sample degradation. The solution was maintained under stirring and exposed to 405 nm while the SOSG PL was monitored using an in-situ fiber with 473 nm excitation. UV exposition was maintained until no further increment of SOSG PL was observed, which indicates the complete oxidation of the SOSG in the solution (**Figure S6**). The ratio between the PL collected at the end (corresponding to complete SOSG oxidation) and before UV irradiation (22.3 times higher) was used to evaluate the fraction of total SOSG moles which are oxidized and thus the ROS moles produced:

$$m_{ROS}(t) = \frac{PL_{SOSG}(t)}{PL_{SOSG}(0)} \frac{1}{22.3} 8.3 \times 10^{-5} M \times V$$

Where PL$_{SOSG}$ (t) is the PL of the SOSG evaluated after a specific exposure time t, $8.3 \times 10^{-5}$ M is the molarity of the SOSG solution and V is the volume of the solution.

|  | Atomic Fraction (%) | | | | | | | | | |
|---|---|---|---|---|---|---|---|---|---|---|
|  | 100nm | | 200nm L | | 200nm M | | 200nm H | | 350nm | |
|  | Avg | SD | Avg | SD | Avg | SD | Avg | SD | Avg | SD |
| O | 65.21 | 1.32 | 62.58 | 1.11 | 61.32 | 1.17 | 60.43 | 1.62 | 63.93 | 1.45 |
| Si | 31.84 | 1.33 | 33.40 | 0.48 | 32.10 | 1.13 | 30.77 | 1.34 | 34.16 | 1.14 |
| Cl | 0.22 | 0.07 | 0.11 | 0.10 | 0.55 | 0.24 | 0.68 | 0.15 | 0.06 | 0.06 |
| K | 0.36 | 0.12 | 0.43 | 0.07 | 0.10 | 0.03 | 0.05 | 0.03 | 0.00 | 0.01 |
| Br | 1.43 | 0.46 | 2.15 | 0.33 | 3.52 | 0.83 | 4.64 | 0.27 | 1.07 | 0.34 |
| Cs | 0.45 | 0.12 | 0.68 | 0.13 | 1.25 | 0.29 | 1.83 | 0.11 | 0.41 | 0.11 |
| Pb | 0.49 | 0.13 | 0.64 | 0.11 | 1.17 | 0.24 | 1.59 | 0.09 | 0.37 | 0.09 |
| LHP | 2.37 | 0.69 | 3.48 | 0.56 | 5.93 | 1.35 | 8.06 | 0.42 | 1.85 | 0.53 |

**Table S1.** EDX analysis on NC-MSN sample set as atomic fraction, following the same color code as in main report.

**Estimation of CsPbBr$_3$ Nanocrystal Content from EDX Atomic Fractions**

To estimate the density $\rho_{NC}$ of CsPbBr$_3$ NCs embedded within the MSN, we utilized the elemental atomic fractions obtained from energy-dispersive X-ray spectroscopy (EDX). The following procedure outlines the calculation based on compositional and structural assumptions:

The atomic fractions of Si, O, Cs, Pb, and Br, obtained from EDX, are denoted as $\eta_{Si}$, $\eta_O$, $\eta_{Cs}$, $\eta_{Pb}$, and $\eta_{Br}$, respectively. The total atomic fraction satisfies:

$$\eta_{Si} + \eta_O + \eta_{Cs} + \eta_{Pb} + \eta_{Br} = 100\%$$

To estimate the molar content of SiO$_2$ and CsPbBr$_3$, we account for the number of atoms per formula unit: three atoms for SiO$_2$ and five atoms for CsPbBr$_3$. The number of moles of each species is approximated as:

$$n_{SiO2} = \frac{\eta_{Si} + \eta_O}{3}; \quad n_{CsPbBr_3} = \frac{\eta_{Cs} + \eta_{Pb} + \eta_{Br}}{5}$$

The molar fraction of CsPbBr$_3$ NC in NC-MSN is given by:

$$\chi_{SiO_2} = \frac{n_{SiO2}}{n_{SiO2} + n_{CsPbBr_3}}; \quad \chi_{CsPbBr_3} = \frac{n_{CsPbBr_3}}{n_{SiO2} + n_{CsPbBr_3}}$$

The mass contributions of each component are calculated using their respective molar masses:
- $M_{SiO_2}$: 60.08 g/mol
- $M_{CsPbBr_3}$: 528.1 g/mol

The total mass of the mixture is expressed as:

$$m_{total} = n_{SiO_2} \cdot M_{SiO_2} + n_{CsPbBr_3} \cdot M_{CsPbBr_3}$$

The mass of CsPbBr$_3$ is then:

$$m_{CsPbBr_3} = x_{CsPbBr_3} \cdot m_{total}$$

The bulk density of CsPbBr$_3$ is 4.86 g/cm$^3$, so the total volume of embedded NC is estimated as:

$$V_{CsPbBr_3} = \frac{m_{CsPbBr_3}}{4.86 g/cm^3}$$

Assuming monodisperse cubic CsPbBr$_3$ nanocrystals with edge length of 8 nm, the volume of a single nanocrystal is:

$$V_{NC} = (8 \times 10^{-7} cm)^3 = 5.12 \times 10^{-19} cm^3$$

Finally, the total number of nanocrystals is estimated by:

$$\rho = \frac{V_{CsPbBr_3}}{V_{NC}}$$

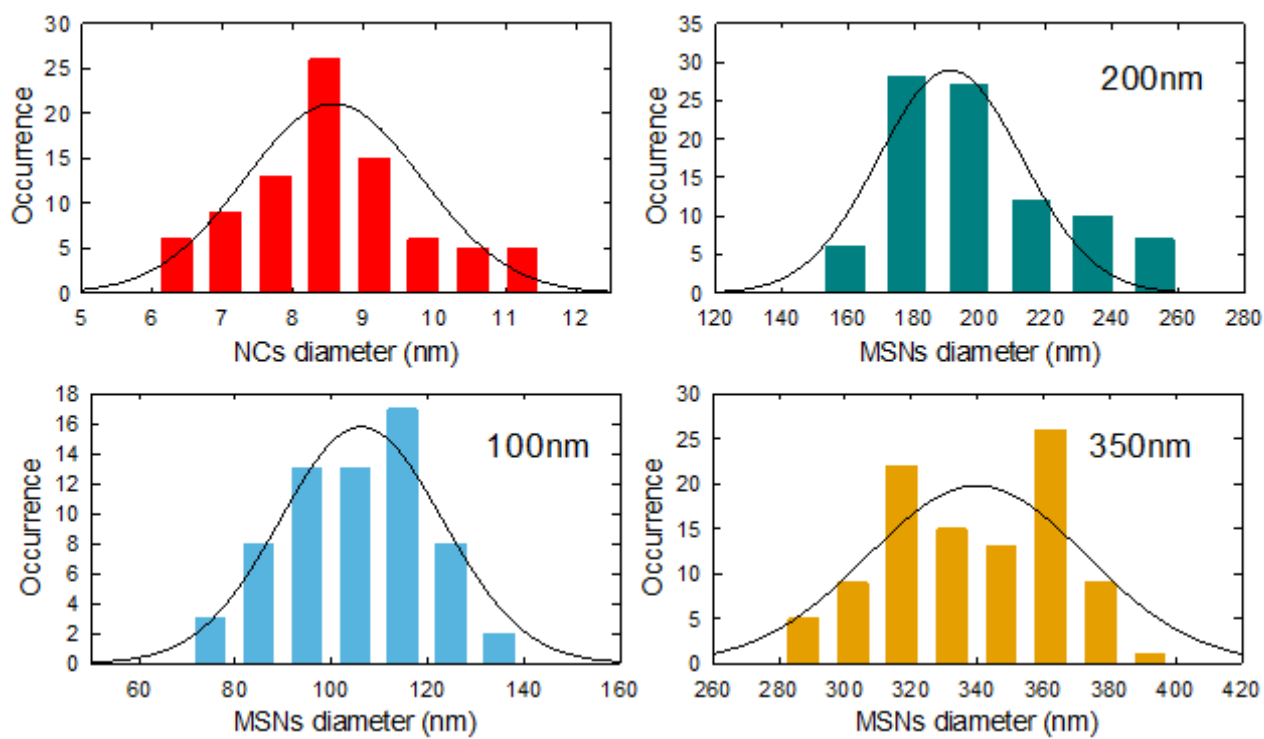

**Figure S1** Size distribution of NC inside MSN and size distribution of MSNs.

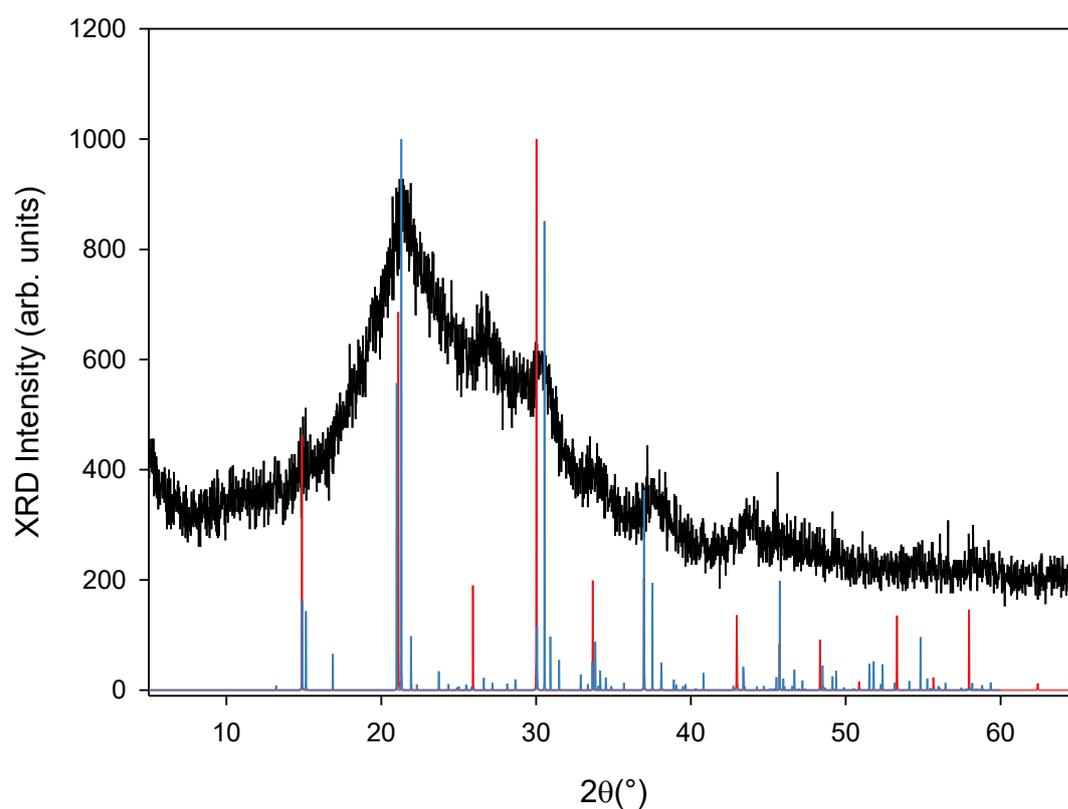

**Figure S2** Representative XRD intensity of NC-MSN (colored line) with d=200 nm together with the pattern for cubic (red line) and orthorhombic $CsPbBr_3$ (blue line).

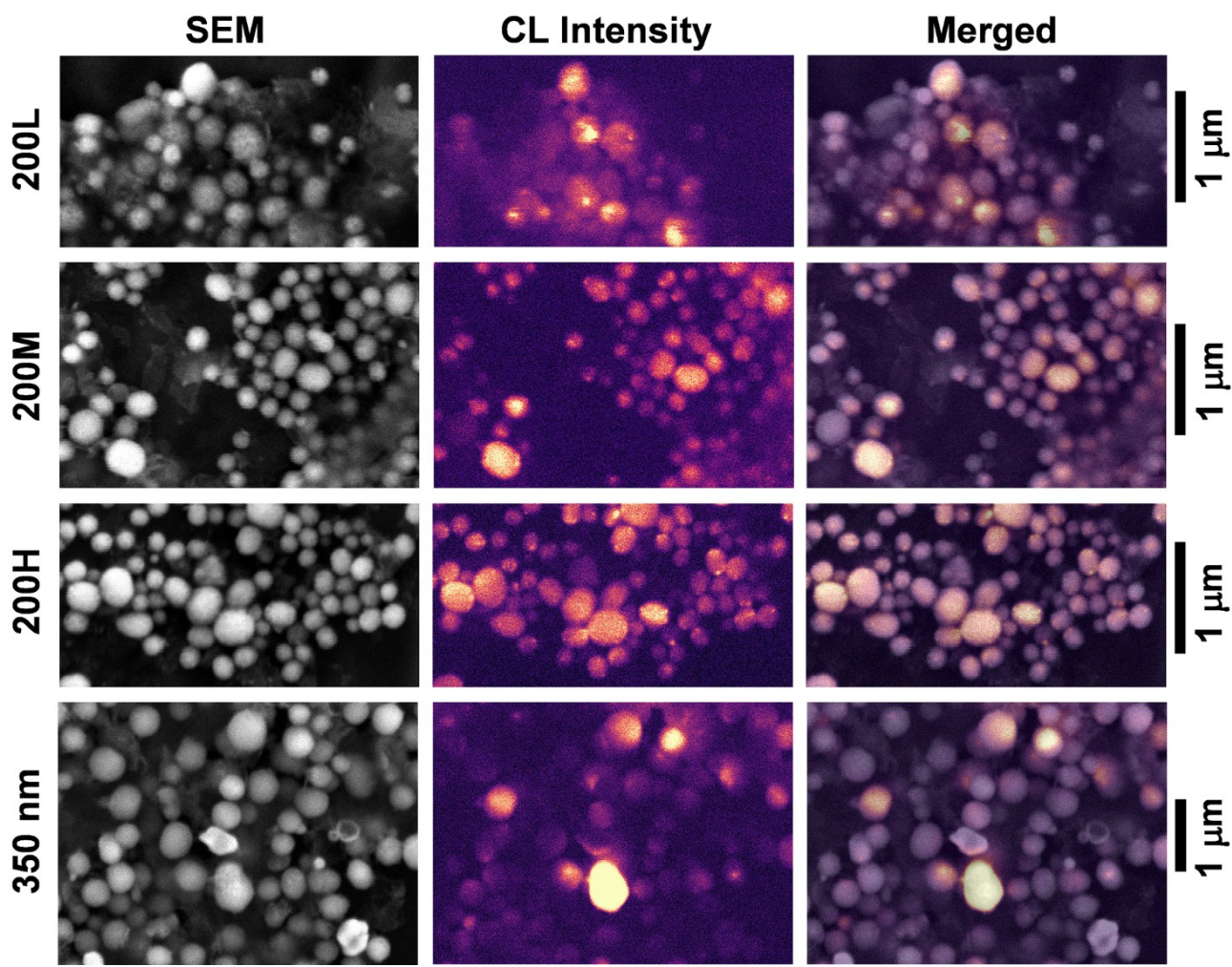

**Figure S3.** From left to right: SEM images, CL intensity maps and merged images of NC-MSNs samples with different diameters

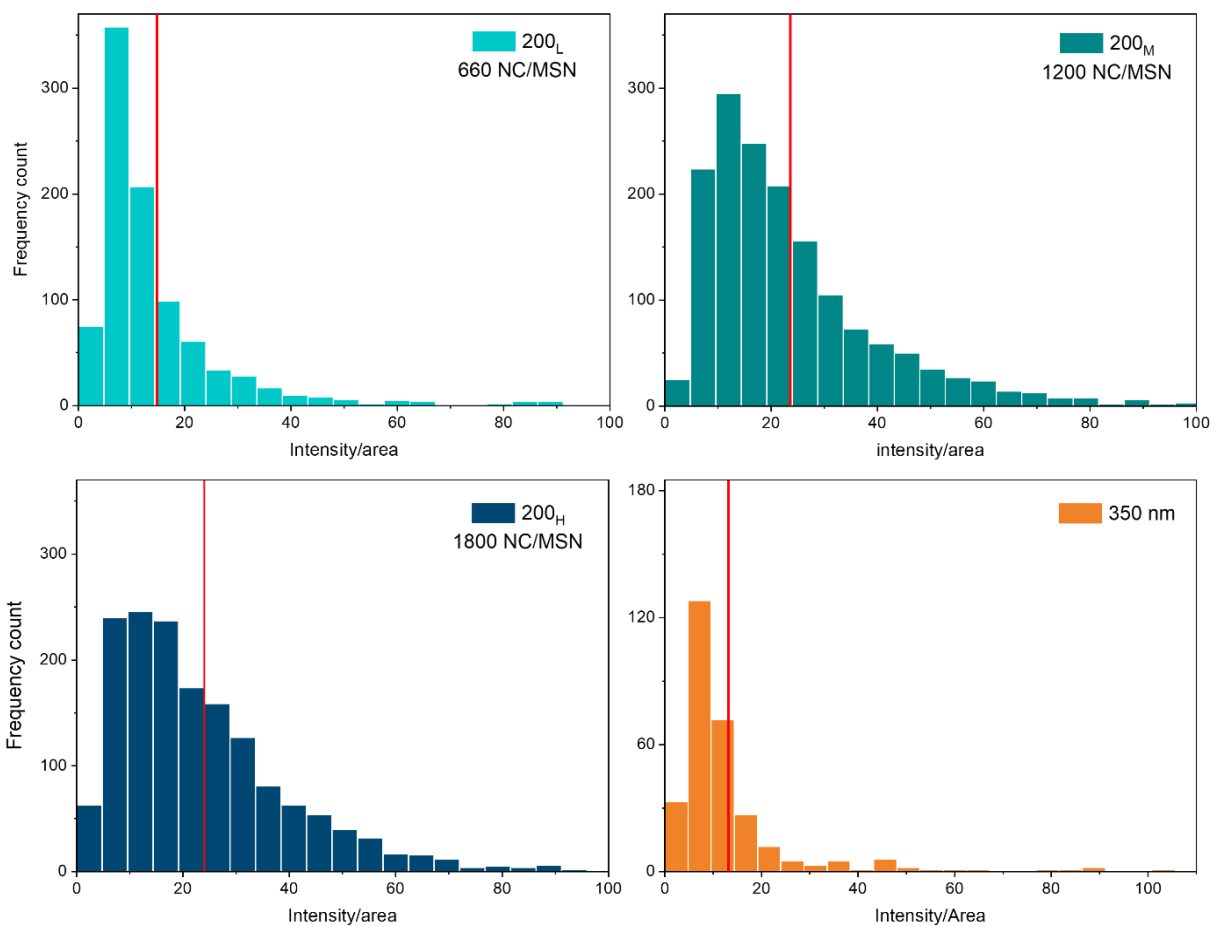

**Figure S4.** Statistical distribution of intensity per unit area obtained from CL maps. Red lines indicate the mean values. Those are $14.8 \pm 0.5$, $23.6 \pm 0.4$, $24.0 \pm 0.4$ and $13.2 \pm 0.8$ (a.u.), respectively.

Quantitative analyses were also performed to obtain a statistical distribution of the intensities emitted by individual spheres.
This analysis was carried out using ImageJ software, working on .tiff SE images and corresponding CL intensity maps. SE images were used to accurately define regions of interest (ROIs), each having a circular shape corresponding to the edge of a single MSN. The defined ROIs were then transferred onto the CL maps, allowing the measurement of both intensity and area for each MSN. Histograms showing the intensity per unit area of the analysed particles are reported in **Figure S4**.
The increase in luminescence observed when increasing the NC concentration from 660 to 1200 NCs/MSN is not further enhanced in the 200H sample. This behaviour is likely due to reabsorption effects occurring at high NC densities. Additional measurements varying the accelerating voltage and beam current may help to further investigate this phenomenon.

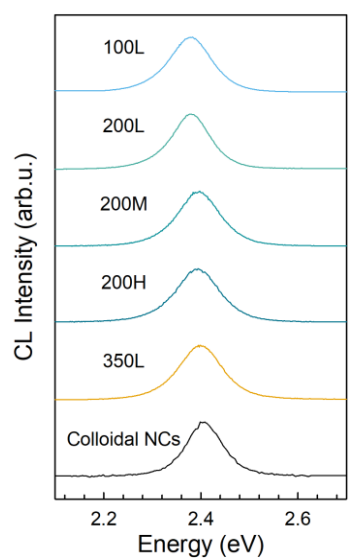

**Figure S5** Cathodoluminescence spectra of CsPbBr3 NCs are related NC-MSNs.

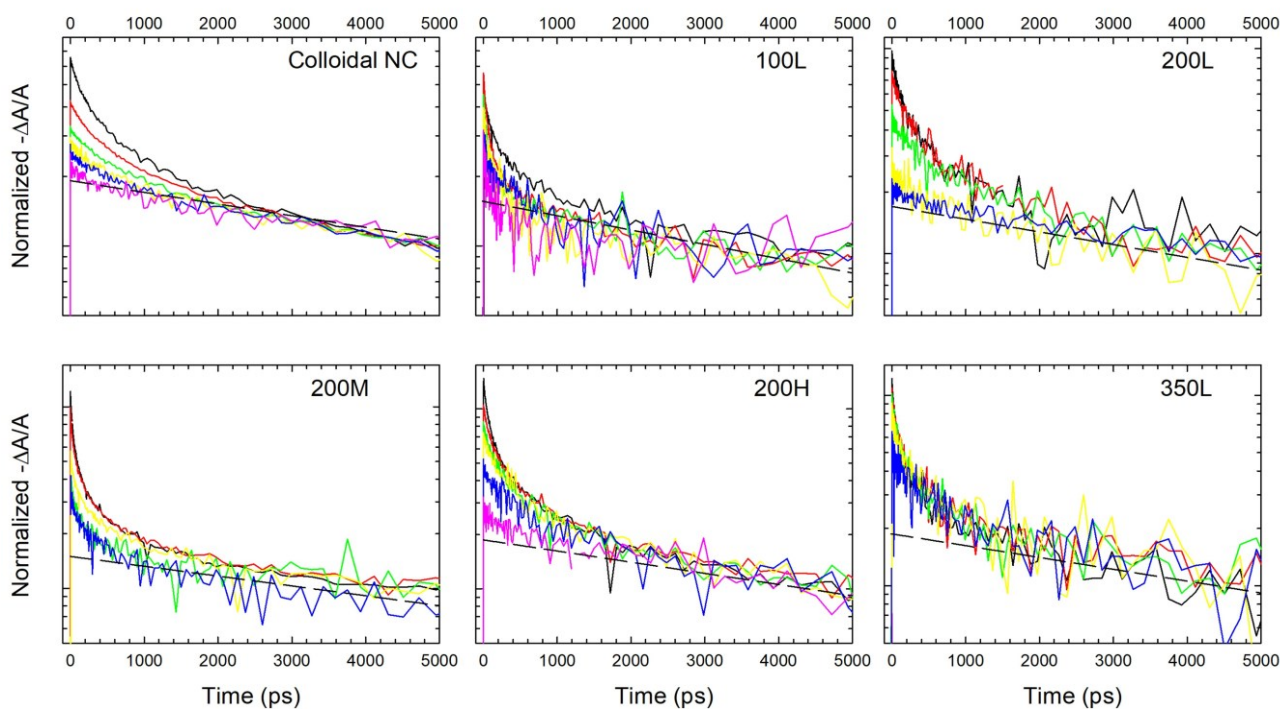

**Figure S6**. Transient absorption dynamics of colloidal NC and NC-MSNs under increasing excitation power (from $\langle N \rangle < 0.4$ to $\langle N \rangle \sim 6$) at 3.05 eV. The lines shade from blue to black as the fluence increases. Dashed line marked the single exciton contribution used in subtraction fitting method.

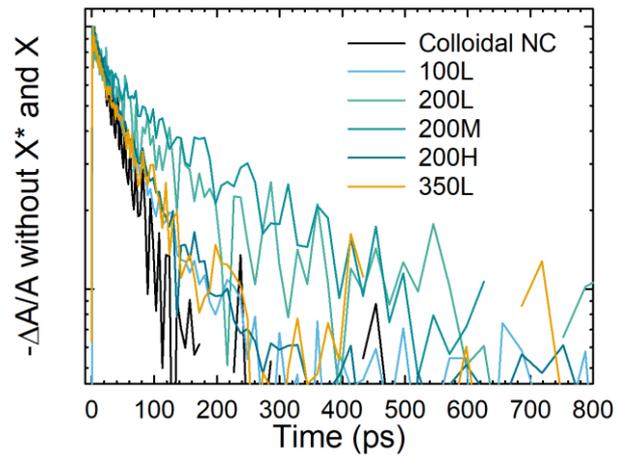

**Figure S7** TA dynamics of *XX*. The fitting results of XX are used to calculate $\Phi_{XX}$ in Figure 1e.

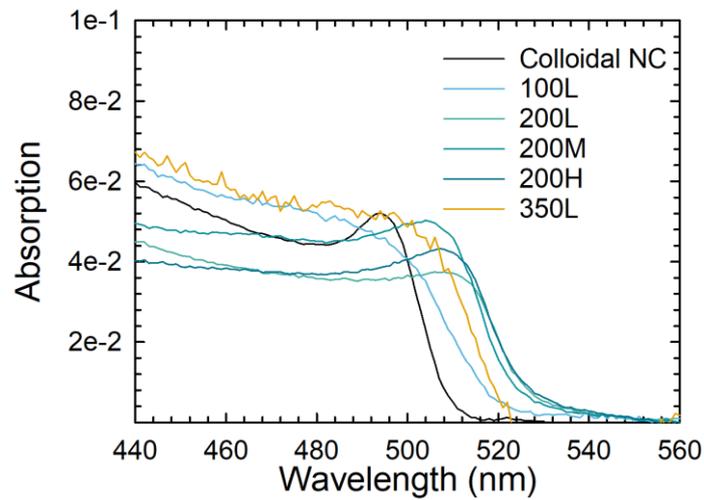

**Figure S8** Absorption spectra of sample solutions for RL measurements with the same average concentration of $CsPbBr_3$ NCs (implied by the same band edge absorption difference).

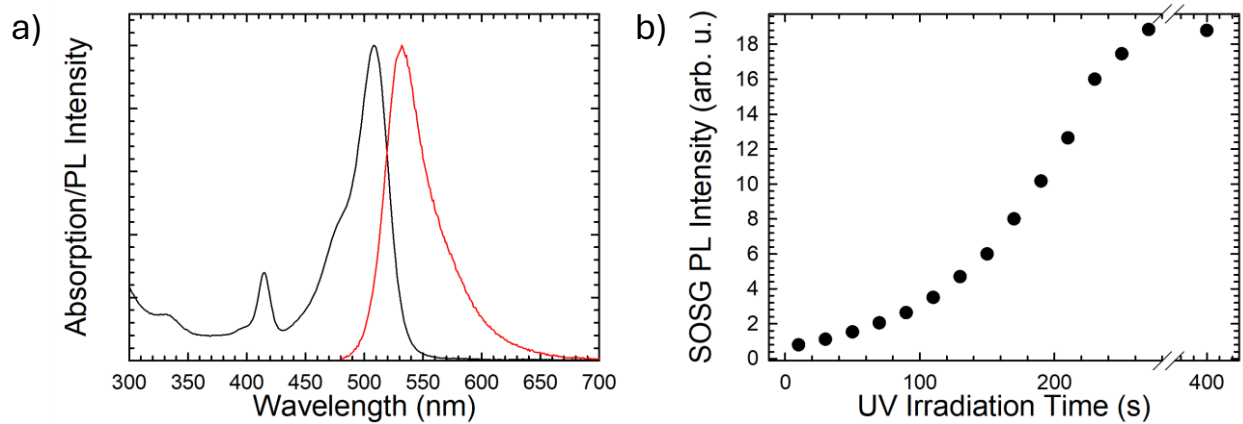

**Figure S9**. a) Absorption (black) and PL emission (red) spectra of SOSG. b) SOSG PL intensity growth using photo-sensitizing approach.

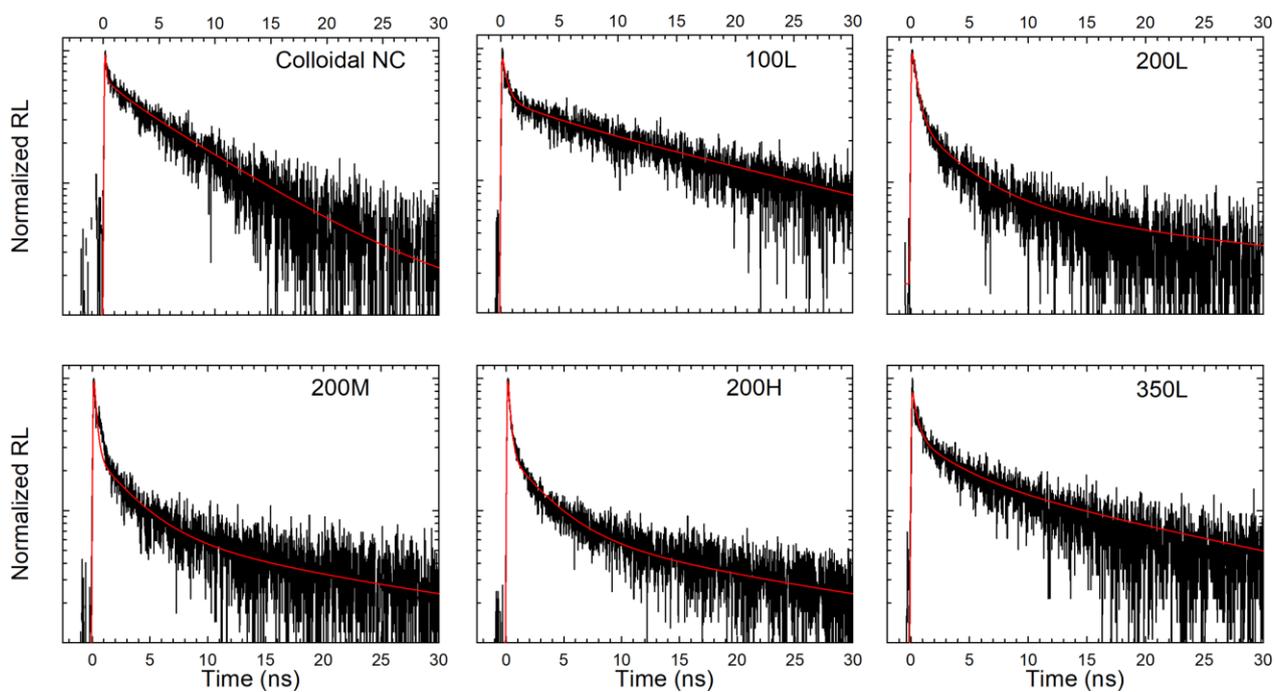

**Figure S10** Time-resolved scintillation decay curves (black) of colloidal NC and NC-MSNs with their corresponding fit lines (red).